# The evolution of the galaxy and the birth of the solar system: The short-lived nuclides connection.

S. Sahijpal

Department of Physics, Panjab University, Chandigarh, India 160 014 (sandeep@pu.ac.in)

**Abstract**

An attempt is made, probably for the first time, to understand the origin of the solar system in context with the evolution of the galaxy as a natural consequence of the birth of several generations of stellar clusters. The galaxy is numerically simulated to deduce the inventories of the short-lived nuclides, $^{26}$Al, $^{36}$Cl, $^{41}$Ca, $^{53}$Mn and $^{60}$Fe, from the stellar nucleosynthetic contributions of the various stellar clusters using an N-body simulation with updated prescriptions of the astrophysical processes. The galaxy is evolved by considering the discreteness associated with the stellar clusters and individual stars. We estimate the steady state abundance of the radionuclides around 4.56 billion years ago at the time of formation of the solar system. Further, we also estimate the present $^{26}$Al/$^{27}$Al and $^{60}$Fe/$^{56}$Fe of the interstellar medium that match within a factor of two with the observed estimates. On contrary to the conventional galactic chemical evolution (GCE) model, the present adopted numerical approach provides a natural framework to understand the astrophysical environment related with the origin of the solar system. We deduce the nature of the two stellar clusters; the one that formed and evolved prior to the solar system formation, and the other within which the solar system probably formed. The former could have contributed the short-lived nuclides $^{129}$I and $^{53}$Mn, whereas, the supernova associated with the most massive star in the latter contributed $^{26}$Al and $^{60}$Fe to the solar system. The analysis was performed with the revised solar metallicity of 0.014.

1. Introduction

The universe probably originated around 13.8 Giga years (Gyr.) subsequent to the big-bang (Ade *et al.* 2013). The hadron and the lepton eras led to the baryogenesis of the fundamental particles (Pagel 1997). The lepton era was followed by the primordial nucleosynthesis era (Pagel 1997) that established the early cosmic abundance of hydrogen and helium in terms of mass to be around 75 % and 25 %, respectively, with no traces of heavier elements except for lithium. The formation of the galaxies commenced in a hierarchical manner from the mergers of the neutral hydrogen clouds subsequent to the recombination era (Chiappini *et al.* 1997; Pagel 1997; Matteucci 2003; Mo *et al.* 2011), perhaps during the initial one Giga years. The onset of the formation of the galaxies initiated the on-going cosmic cycle of star formation and evolution within the galaxies. Stars synthesise the entire range of nuclides by a wide-range of nuclear reactions during the various stages of their evolution (Pagel 1997). Several generations of stars subsequent to their evolution contributed their stellar nucleosynthetic yields to the galactic inventories of the stable and radionuclides. This resulted in a steady state growth of the elemental abundances of the heavier elements beyond helium on account of the galactic chemical evolution (Matteucci & Greggio 1986; Matteucci & François 1989; Timmes *et al.* 1995; Chiappini *et al.* 1997; Pagel 1997; Chang *et al.* 1999; Goswami & Prantzos 2000; Alibés *et al.* 2001; Matteucci 2003; François *et al.* 2004; Kobayashi & Nakasato 2011; Mo *et al.* 2011; Sahijpal & Gupta 2013; Sahijpal 2013, 2014).

The galactic chemical evolution (GCE) deals with the gradual growth of the elemental as well as their isotopic abundance over the galactic timescales by primarily taking into account the history associated with the accretional growth of the galaxy, the star formation rate and the stellar initial mass function at the time of star formation (Matteucci & Greggio 1986; Matteucci & François 1989; Timmes *et al.* 1995; Chiappini *et al.* 1997; Pagel 1997; Chang *et al.* 1999; Goswami & Prantzos 2000; Alibés *et al.* 2001; Matteucci 2003; François *et al.* 2004; Kobayashi & Nakasato 2011; Mo *et al.* 2011; Sahijpal & Gupta 2013; Sahijpal 2013, 2014). In an operational manner, the GCE depends upon the historic rates and the stellar nucleosynthetic contributions of the generations of AGB (asymptotic giant branch) stars and supernovae (types Ia, II and I b/c) (see e.g., Pagel 1997; Chiappini *et al.* 1997; Goswami & Prantzos 2000; Alibés *et al.* 2001; Matteucci 2003; Mo *et al.* 2011; Sahijpal & Gupta 2013). Apart from understanding the entire temporal evolution of the galaxy, one of the major emphases of developing the GCE models is to understand the origin of the stable isotopic abundance of the solar system (Anders & Grevesse 1989; Asplund *et al.* 2009). The solar system formed around 4.56 Gyr. ago by the gravitational collapse of a molecular cloud. The majority of the GCE models involve a framework of the numerical solutions of the complex integro-differential equations (Matteucci & Greggio 1986; Matteucci & François 1989; Timmes *et al.* 1995; Chiappini *et al.* 1997; Pagel 1997; Chang *et al.* 1999; Goswami & Prantzos 2000; Alibés *et al.* 2001; Matteucci 2003; François *et al.* 2004) that deals with the gradual growth of the stable isotopic abundance of all the elements from hydrogen to zinc. This classical approach is based on an averaged temporal behaviour of the various episodes of star formation and evolution. The approach has limitations as it cannot appropriately take into account the discrete nature of the star forming episodes. This becomes even more relevant in the case of the star formation in the earliest epochs of the galaxy that were related with the population III stars (Sahijpal 2013), and in the present problem, the specific environment associated with the stellar clusters that were related with the origin of the solar system. The understanding of the nature of these stellar clusters would be difficult to perceive with the conventional GCE models. An approach based on N-body numerical simulation would be more relevant as it can consider the discreteness associated with the formation of stars in the stellar clusters.

It should be mentioned that the chemodynamical models based on the Smoothed Particle Hydrodynamical (SPH) method provides a better approach to understand the GCE in terms of the dynamics of the evolving galaxy (Kobayashi & Nakasato 2011). This approach takes into account the dynamics associated with the formation and evolution of the galaxy. This approach has the potential to eventually explain most of the observational aspects related with the evolution of the galaxy. As far as the numerical framework is concerned, leaving apart the dynamical issues, the present adopted numerical approach has much more similarities with the chemodynamical models rather than the conventional approach to understand the galactic chemical evolution. Nonetheless, our recent attempts in this regard would also provide a better understanding for the evolution of the galaxy (Sahijpal 2013).

Recently, a GCE model has been proposed based on the N-body simulations of the galaxy in an almost realistic manner (Sahijpal & Gupta 2013). In this approach, the galaxy is evolved by taking into account, in a discrete manner, the birth and evolution of several generations of stars as individual entities. Within a simulation, the modelled stars are born. These stars evolve and eventually eject their nucleosynthetic debris into the interstellar medium (ISM). On contrary to the conventional approach this is a more direct approach to understand the evolution of the galaxy. This model offers a stochastic approach of treating the birth and death of individual stars. Apart from the comparison of the predicted elemental (isotopic) abundance distribution of the solar system at the time of its formation with the observations, a GCE model should also explain the elemental abundance evolution over the galactic timescale, the present prevailing star formation rate, the supernova rates, the surface mass densities of the total matter, stars and interstellar gas. The attempts made by several groups basically aim at achieving the best possible match with the observations with a suitable choice of the parameters. A detailed discussion regarding these aspects along with the parametric analyses of the various N-body simulations has been presented (Sahijpal & Gupta 2013). Some of the parameters related with the two stage accretion scenario of the galaxy, the star formation rate and the stellar initial mass function were directly chosen from the earlier works. Within a suitable choice of parameter that are consistent with the earlier works on the convention GCE approach, the GCE model based on the N-body simulation was successful in explaining most of the observational features related the galactic chemical evolution including the origin of the isotopic abundances of most all the stable elements from hydrogen to gallium of the solar system (Sahijpal & Gupta 2013) within a factor of 2 as is generally observed in the various GCE models based on the conventional approach (see e.g., Timmes *et al.* 1995; Alibés *et al.* 2001; Matteucci 2003). A wide range of uncertainties related with the assumptions dealing with the accretion scenario of the galaxy, the star formation rate, the stellar initial mass function, the stellar evolution and nucleosynthesis propagate in determining the uncertainties related with the final abundance distribution of elements (isotopes).

The solar abundance of the various elements have been recently revised from their earlier estimates (Anders & Grevesse 1989) to the present estimates (Asplund *et al.* 2009) on account of the various refinements in the spectroscopic techniques and abundance determination procedure. As a result of this revision, the metallicity (Z; the total mass fraction of all the elements beyond boron) of the sun has been reduced to 0.014 (Asplund *et al.* 2009) from an initially accepted value of ~0.02 (Anders & Grevesse 1989). All the conventional GCE models were based on the earlier solar abundance estimates. The recent GCE model (Sahijpal & Gupta 2013) has been able to reproduce the present revised estimates of all the elements within uncertainties as discussed above. Further, a comparison was presented among the GCE models based on the pre-revised and the revised solar abundances. It should be noted that the revision of the solar elemental abundances has not only brought significant changes in the GCE predictions in terms of the supernova history (Sahijpal & Gupta 2013) but has also influenced the helioseismological constraints on the standard solar model predictions (see e.g. Bi *et al.* 2011). In this context it is quite possible that the further revision in revised solar elemental abundance distribution, specifically, for the elements like neon (Asplund *et al.* 2009) could further alter the inferred metallicity, and hence, its influence on GCE and standard solar model predictions. Among all the elements that define the metallicity of ~0.014, oxygen, carbon and neon contributes, 43.5%, 17.8% and 7.2%, respectively, as the three major elements. This is due to the higher supernova SN II + I b/c yields of these three elements (see e.g., fig. 3 of Sahijpal 2014). Due to the uncertainties in the estimates of neon abundance (Asplund *et al.* 2009), a change in the metallicity would be inevitable. In case we use the observed estimates of A(Ne) = 8.11 for the nearby B stars, instead of the presently used value of A(Ne) = 7.84 (Asplund *et al.* 2009) for the neon abundance, the metallicity increases from ~0.014 to ~0.015, a 6% increase. The percentage contribution of neon goes to 12.5% from 7.2%. Thus, we anticipate a revised range of metallicity of ~0.014 - 0.015. This would result in a proportionate enhancement in the supernova SN II+ I b/c activity of the galaxy.

In the present work, we infer the GCE of the short-lived nuclides, $^{26}$Al ($\tau \sim$ 1.05 Million years; Myr.), $^{36}$Cl ($\tau \sim$ 0.43 Myr.), $^{41}$Ca ($\tau \sim$ 0.15 Myr.), $^{53}$Mn ($\tau \sim$ 5.34 Myr.) and $^{60}$Fe ($\tau \sim$ 3.75 Myr.) in order to understand the steady state contribution of these radionuclides to the early solar system with the revised elemental abundances (Asplund *et al.* 2009). We deduce the GCE of these isotopes based on the N-body numerical simulation that takes into account the discreteness of every star forming episode (Sahijpal & Gupta 2013). Further, as a major part of this work we make an attempt, probably for the first time, to understand the nature of the stellar clusters that were directly relevant for the formation of the solar system on the basis of the entire evolution of the galaxy. We specifically focus on the last two episodes of the star formation in the "local" ISM (Sahijpal 2013) associated with the solar system formation in order to understand the source of most of the short-lived nuclides and the nature of the stellar clusters.

The presence of the short-lived nuclides, $^{26}$Al, $^{36}$Cl, $^{41}$Ca, $^{53}$Mn and $^{60}$Fe, in the early solar system has been inferred on the basis of the enhancement in their daughter nuclide abundances in the early solar system phases that are found in some primitive meteorites (see e.g., MacPherson *et al.* 1995; Sahijpal & Soni 2006; Wasserburg *et al.* 2006; Gaidos *et al.* 2009; Huss *et al.* 2009; Sahijpal & Gupta 2009; Krot *et al.* 2012). The presence of these radionuclides demands either the stellar origin for most of them just prior to the initiation of the formation of the solar system, or by the irradiation of the solar nebula gas and dust by the energetic particles from the active early sun going through the T Tauri phase. Based on the detailed numerical simulations of the irradiation scenario, it has been argued that in order to explain the empirically observed narrow spread in the initial $^{26}$Al/$^{27}$Al in the early solar system, the irradiation scenario seems to be the most unlikely source of at least $^{26}$Al (Sahijpal & Soni 2007; Sahijpal & Gupta 2009). This leaves the stellar source as the most likely source of at least $^{26}$Al, the most well studied short-lived nuclide (MacPherson *et al.* 1995). Among the various proposed stellar sources (MacPherson *et al.* 1995; Sahijpal & Soni 2006; Wasserburg *et al.* 2006; Sahijpal & Gupta 2009; Gaidos *et al.* 2009; Huss *et al.* 2009; Krot *et al.* 2012), it is most likely that a massive star might have produced this short-lived nuclide along with the other short-lived nuclides (Gaidos *et al.* 2009; Sahijpal & Gupta 2009). Further, it has been argued that this massive star might have gone through a Wolf-Rayet phase and ultimately exploded as supernova SN Ib/c (Gaidos *et al.* 2009; Sahijpal & Gupta 2009). Thus, the formation of the solar system either occurred in contemporary with the formation of the massive star in a massive stellar cluster (Sahijpal & Gupta 2009; Sahijpal 2013), or, the massive star ejected the radionuclides into the protosolar molecular cloud and triggered the formation of the solar system (MacPherson *et al.* 1995; Gaidos *et al.* 2009; Huss *et al.* 2009; Krot *et al.* 2012). In the former scenario the massive star ejected the short-lived nuclides into the protoplanetary disc of the already forming solar system (Sahijpal & Gupta 2009). In order to understand the origin of the solar system it becomes extremely important to decipher the source(s) of all the discovered short-lived nuclides (Gaidos *et al.* 2009; Huss *et al.* 2009). Further, the extent of the presence of the short-lived nuclides, $^{26}$Al and $^{60}$Fe, in the early solar system also directly influence the early thermal evolution of the planetesimals (Sahijpal *et al.* 2007), the building blocks of the planets. The heat generated by the decay of these short-lived nuclides within the early accreted planetesimals is considered to have melted and differentiated the planetesimals and asteroids (e.g., Sahijpal *et al.* 2007). The radioactive heating is also considered to have caused wide-range of aqueous alteration, and the early thermal evolution of the icy planetesimals and icy satellites of the giant planets.

As mentioned earlier, in the present work we have numerically simulated the GCE of the short-lived nuclides and further make an attempt exclusively on the basis of our N-body numerical approach of the GCE to decipher the nature of the stellar clusters (Sahijpal 2013) associated with the birth of the solar system. On contrary to using the classical approach of solving the analytical equations dealing with GCE (Clayton 1985; Huss & Meyer 2009; Meyer & Huss 2009) that is based on an averaged temporal behaviour of the galactic evolution, we have performed N-body simulation that is based on the stochastic approach of treating the birth and death of individual stars within the stellar cluster (or associates stellar clusters). This approach becomes specifically more useful in properly taking into account the stellar contributions of the stellar clusters that were directly related with the birth of the solar system. In the present work, we specifically try to understand the nature of the stellar cluster (or associated stellar clusters) that was associated with the formation of the solar system, and the stellar cluster that evolved prior to the formation of the solar system (Gaidos *et al.* 2009; Huss *et al.* 2009; Krot *et al.* 2012). Further, on contrary to the previous work on the GCE of the short-lived nuclides (Huss & Meyer 2009; Meyer & Huss 2009) we have used the updated prescriptions (Chiappini *et al.* 1997; Matteucci 2003; Sahijpal & Gupta 2013) for the formulations associated with the formation of the galaxy, the star formation rates, the stellar initial mass function and the revised solar metallicity (Asplund *et al.* 2009). The previous works (Huss &

Meyer 2009; Meyer & Huss 2009) were based on the GCE model developed much earlier (Clayton 1985). These models are based on the steady single infall of matter on the galaxy contrary to the recent suggestions that treat the accretion of galaxy in two episodes (Chiappini *et al.* 1997; Matteucci 2003; Sahijpal & Gupta 2013). Further, the star formation rate was assumed to depend linearly on the galactic gas density, whereas, the recent models (Chiappini *et al.* 1997; Matteucci 2003; Sahijpal & Gupta 2013) are based on much complex star formation rate scenarios. In addition, apart from the incorporation of the stellar contributions of the short-lived nuclides from supernovae SN II and SN Ib/c (Woosley & Weaver 1995) in the GCE model for the short-lived nuclides (Huss & Meyer 2009; Meyer & Huss 2009), we have also included the contributions of supernovae SN Ia (Iwamoto *et al.* 1997) for all the relevant short-lived nuclides. We could also incorporate the stellar contributions of $^{26}$Al from the AGB stars (Karakas 2010). Finally, we estimate the contributions of the short-lived nuclides to the early solar system from the GCE, and the stellar clusters that were immediately related with the origin of the solar system.

## 2. Methodology

We numerically simulated the evolution of the milky-way galaxy as it gradually grows due to the exponentially declining in-fall of the extra-galactic matter (Chiappini *et al.* 1997; Goswami & Prantzos 2000; Alibés *et al.* 2001; Matteucci 2003; Sahijpal & Gupta 2013). We assumed a two stage accretion phase for the formation of the galaxy in order to avoid the G-dwarf metallicity distribution problem (Chiappini *et al.* 1997; Alibés *et al.* 2001; Matteucci 2003; Sahijpal & Gupta 2013). The galactic halo and the thick disk of the galaxy were assumed to accrete with an exponentially declining in-fall rate (Chiappini *et al.* 1997; Alibés *et al.* 2001; Matteucci 2003; Sahijpal & Gupta 2013) over a timescale of 1 Giga years. The thin disk of the galaxy, in the solar neighbourhood, was assumed to accrete matter from the in-falling gases with an exponentially declining in-fall rate (Chiappini *et al.* 1997; Alibés *et al.* 2001; Matteucci 2003; Sahijpal & Gupta 2013) over a timescale of 7 Giga years. This accretion scenario eventually results in the present day observed densities of ~54 and ~10 $M_\odot$ pc$^{-2}$, for the total local thin disk and the total local thick disk in the solar neighborhood. The metallicity of the in-falling gases during the accretion was assumed to be 0.1 times the assumed final metallicity of the solar system (Alibés *et al.* 2001; Sahijpal & Gupta 2013). This assumption is helpful in avoiding the G-dwarf metallicity distribution problem (Matteucci 2003) and is supported by the empirical observations (Alibés *et al.* 2001).

The N-body numerical simulations for the present GCE model are based on treating the birth of stars within a stellar cluster in a discrete manner (Sahijpal & Gupta 2013). The simulations are performed for the solar neighbourhood that is considered to be an annular ring spanning over an assumed galactic radius of 7-8.9 kilo-parsecs during the timescales of 13 Giga years (Gyr.) and with an assumed time step of one million years (Sahijpal & Gupta 2013). An individual massive stellar cluster (or associated stellar clusters) was assumed to form at a regular time interval of one million years throughout the evolution of the solar annular ring according to the prevailing star formation rate at a specific time. The prevailing star formation rate will determine the total mass of the stellar cluster (or associated stellar cluster) (Sahijpal 2013). The chemodynamical GCE models also deal with the formation of stellar clusters according to the prevailing star formation rate throughout the evolution of the galaxy. The size distribution of the stars in the stellar cluster in the mass range 0.1-100 $M_\odot$ is defined by the choice of the stellar initial mass function (Chiappini *et al.* 1997; Pagel 1997; Matteucci 2003; Sahijpal & Gupta 2013). The assumed stellar initial mass function (eq. 1) consists of stars with discrete set of masses. In the low mass range, we choose the stars with the masses, 0.1, 0.8, 1.0, 1.25, 1.75, 2.5, 3 $M_\odot$ to adequately cover the low mass region. This choice was basically governed by the availability of the stellar nucleosynthetic yields of the AGB stars corresponding to some of these masses (Karakas 2010). The mass range of 4-100 $M_\odot$, corresponding to the intermediate and massive stars, was covered by the stars with the masses corresponding to the integer numbers, except for the masses, 9 and 10 $M_\odot$, for which there are uncertainties in the stellar evolutionary theories.

$$\varphi_n(m) = A\, m^{-(1+x)} \tag{1}$$

The number distribution '$\varphi_n(m)$' of the stars of a specific mass 'm' defines the number of corresponding stars produced within a stellar cluster. The integral, over the entire mass distribution (0.1-100 $M_\odot$), of the product of the number distribution '$\varphi_n(m)$' with the specific mass, 'm', will define the total mass of the stellar cluster during its formation that in turn will be decided by the star formation rate at any epoch. The choice of the exponent, 'x' is partially based on observational evidences (Chiappini *et al.* 1997; Matteucci 2003), and to an extent decided by their parameterization to eventually obtain the solar metallicity value at the time of formation of the solar system around 4.56 Gyr. ago (see e.g., Chiappini *et al.* 1997; Alibés *et al.* 2001; Matteucci 2003; Sahijpal & Gupta 2013). A compilation regarding the spread in the power exponents can be found by Matteucci (2003). As mentioned earlier, we have developed the GCE models for the short-lived nuclides with the pre-revised (Anders & Grevesse 1989) and the revised (Asplund *et al.* 2009) solar metallicities. The former simulations will allow us to compare our results with the previous works (Huss & Meyer 2009; Meyer & Huss 2009) that were performed for the pre-revised solar metallicity, whereas, the latter will serve as the final set for our discussions. The choice of the exponent adopted for the pre-revised solar metallicity model is 0, 1.7 and 1.36, in the mass ranges, 0.1-1 $M_\odot$, 1.25-8 $M_\odot$ and 11-100 $M_\odot$, respectively. However, in the case of the revised solar metallicity model we choose a value of 1.5, in the mass 11-100 $M_\odot$. The adopted piece-wise approach for defining the exponent is consistent with the other works (Chiappini *et al.* 1997; Alibés *et al.* 2001; Matteucci 2003) and the choice of the power exponents for our two models is well within the compiled ranges (Matteucci 2003). The choice of the parameter, 'A' is time dependent within a simulation as it defines the total number of stars produced within a stellar cluster, and hence defines the total mass of the stellar cluster. This will depend upon the star formation rate (SFR; in $M_\odot$ pc$^{-2}$ Myr$^{-1}$) at any specific time of the evolution of the galaxy. We adopted the updated prescription (Alibés *et al.* 2001; Matteucci 2003; Sahijpal & Gupta 2013) regarding the star formation rate (SFR) that depends upon the prevailing gas density ($\sigma_{Gas}$) and the total density ($\sigma_{Total}$) of the galaxy in the solar neighbourhood (eq. 2).

$$SFR = [(\sigma_{Total})^{0.33} \times (\sigma_{Gas})^{1.67}] / \sigma_{Total} \qquad (2)$$

Thus, the star formation rate is estimated after every one million years interval of the simulations. On the basis of the deduced star formation rate, the stellar clusters are produced. The different stars formed according to the prevailing metallicity of the ISM at any specific time evolve and eventually die according to their lifetime that depends upon their masses and metallicities (Pagel 1997). At the time of their final evolution, the stars eject their nucleosynthetic material into ISM. The next generation of stars are produced from the reprocessed ISM. We appropriately accounted for the evolution of the ensemble of several generations of stars over the galactic timescales. We have used the instant recycling mixing approximation for the GCE as is used by majority of the GCE models (Matteucci & Greggio 1986; Matteucci & François 1989; Timmes *et al.* 1995; Chiappini *et al.* 1997; Pagel 1997; Chang *et al.* 1999; Goswami & Prantzos 2000; Alibés *et al.* 2001; Matteucci 2003; François *et al.* 2004; Sahijpal & Gupta 2013; Sahijpal 2014). According to this approximation, the stellar ejecta from an evolved star is instantaneously mixed over the entire region of the solar annular ring that in the present work is defined within the galactic radius of 7-8.9 kilo-parsecs, an area of $9.49 \times 10^7$ pc$^2$. As discussed in the next section, this approximation cannot hold at least in the case of the short-lived nuclides as these radionuclides will decay rapidly before significant mixing (see e.g., Clayton 1983). An attempt has been made in the present work in this context. The instant mixing approximation could however hold to an extent for the stable nuclides as demonstrated in our recent work on the inhomogeneous galactic chemical evolution (Sahijpal 2013). The heterogeneity in the stable elemental abundance evolution among the various regions reduce significantly around [Fe/H ~ 0] corresponding to the birth of solar system.

The single stars in the mass range 1-8 $M_\odot$ eventually evolve to AGB stars. These stars eject their nucleosynthetic debris into the ISM in the form of planetary nebula, thereby, leaving a white-dwarf as a remnant that does not further participate in GCE except in the case if the remnant has a companion star. The accretion of matter by the white-dwarf from the companion star can result in supernova SN Ia. We considered the metallicity dependent stellar nucleosynthetic yields of the AGB stars (Karakas 2010) for all the stable nuclides upto silicon. The stellar nucleosynthetic yields of $^{26}$Al alone are available for these stars for a wide-range of metallicities. Corresponding to the stellar metallicity values where the stellar nucleosynthetic yields are not available in literature, the yields were obtained by linearly interpolating the available stellar yields as has been routinely done earlier (see e.g., Sahijpal & Gupta 2013, and the references therein). It should be noted that except for the binary systems that can eventually lead to SN Ia, we have not considered the massive star binary systems that can eventually evolve in a distinct manner, e.g., the stellar mergers leading to SN I b/c (see e.g., Sahijpal & Soni 2006). This is due to the limited understanding of the evolution and nucleosynthetic yields of such systems. The majority of the GCE models exclude such evolutionary possibilities. This is one of the limitations associated with majority of the GCE models.

The stars in the mass range 11-100 $M_\odot$ eventually explode as supernova SN II and SN Ib/c (Huss *et al.* 2009; Woosley & Weaver 1995). These stars eventually eject their nucleosynthetic debris in ISM in the form of a violent event that can cause disruption of the nearby pre-existing giant molecular clouds (GMC) or smaller neutral hydrogen clouds. Subsequent to the explosion of these stars, their stellar remnants survive in the form of black holes or neutron stars that do not further participate in GCE. We have considered the existing metallicity dependent stellar nucleosynthetic yields of supernovae (Woosley & Weaver 1995) of stars in the mass range 11-40 $M_\odot$. We extrapolated and/or interpolated these stellar nucleosynthetic yields to obtain the stellar yields for the remaining stars with different metallicities and masses where ever the stellar yields are not available in literature. The stellar nucleosynthetic yields of the 30, 35 and 40 $M_\odot$ stars (Woosley & Weaver 1995) corresponding to the models with different remnant masses were averaged out to take into account the equally weighted contributions from the various models. The nucleosynthetic yields of the stable isotopes of iron from SN II and SN Ib/c were systematically reduced by a factor of two in order to appropriately take into account the contributions of iron nuclides from SN Ia, SN II and SN Ib/c (Timmes *et al.* 1995; Alibés *et al.* 2001; François *et al.* 2004). This approach is now routinely followed. We do not follow the approach of suitably modulating the stellar nucleosynthetic yields (François *et al.* 2004) in order to explain the observed trends in GCE that is otherwise considered fair enough given the uncertainties in the stellar nucleosynthetic yields.

The binary systems having the total mass in the range of 3-16 $M_\odot$ experience supernova SN Ia on account of the accretion of the matter from the companion star by the white-dwarf (Matteucci 2003). The white-dwarf explodes as its mass exceeds the Chandrashekhar mass limit. The SN Ia are considered as the major contributor of the iron-peaked nuclides. In our formulation, a fraction, 'f' of the stars generated in the mass interval 3-16 $M_\odot$ were evolved as binary stars that eventually produce a SN Ia (Sahijpal & Gupta 2013). The remaining fraction of the stars was evolved as single intermediate or massive stars. The fraction, 'f', was considered as one of the simulation's free parameters to obtain the observed solar abundance of iron at the time of the formation of the solar system around 4.56 Gyr. ago. In most of the optimized simulations, the fraction, 'f' is around 0.05. This fraction determines the SN Ia rate. The present day predicted SN Ia rate should match with the observed SN I rates (2-3×$10^{-6}$ $pc^{-2}$ $Myr.^{-1}$; Alibés et al. 2001). This matches in the present work as presented in the results section. This aspect has been discussed rigorously by Sahijpal and Gupta (2013). A comparison of the present day predicted SN Ia was made among the various GCE models and the observations (Sahijpal & Gupta 2013). The observed present day SN Ia rates are obtained by different researchers with a wide range of values of 0.007 (Timmes *et al.* 1995) to 0.04 (Alibés *et al.* 2001) for the pre-revised solar models. Our values match with the estimates of Alibés et al. (2001) probably due to the similarities in the approach regarding the accretion of the galaxy and the star formation rate formulation. We adopted these two formulations on the basis of the conventional GCE model developed by Alibés et al. (2001). The stellar nucleosynthetic yields of the various models for SN Ia (Iwamoto *et al.* 2009) were used for the stable as well as the short-lived nuclides. There is one limitation with this approach as these stellar yields are for the solar metallicity stars alone. The metallicity dependent yields for SN Ia are not yet available in literature.

### 3. Results & Discussion

A wide-range of simulations for the GCE of the short-lived nuclides were performed with varied parameters. We are presenting here the results of only the best optimized simulations that were able to successfully explain the solar metallicity and the solar iron abundance at the time of the formation of the solar system around 4.56 Gyr. ago. This corresponds to a time epoch of ~8.5 Gyr. from the initiation of the formation of the galaxy. As mentioned earlier, the simulations were performed for the pre-revised solar metallicity of 0.02 (Anders & Grevesse 1989) and the revised solar metallicity of 0.014 (Asplund *et al.* 2009). The former simulation will provide a comparison with the previous works (Huss & Meyer 2009; Meyer & Huss 2009) that were performed for the pre-revised solar metallicity. These works were based on the averaged temporal behaviour of the GCE model that was developed much earlier (Clayton 1985). The essential results regarding the deduced trends in the various parameters related with the present GCE model corresponding to the revised solar metallicity in the solar neighbourhood is presented in Fig. 1. This includes the deduced temporal evolution of the star formation rate (SFR) (Fig. 1a); the deduced total density ($\sigma_{Total}$), the gas density ($\sigma_{Gas}$) and the (stars+remnant) density ($\sigma_{Stars+remnant}$) (Fig. 1b); the deduced supernovae rates (Fig. 1c); the deduced stellar remnant densities in the form of white-dwarfs, black holes and neutron stars (Fig. 1d); the deduced temporal evolution of the metallicity (Fig. 1e) and the deduced [Fe/H] trend in the solar neighbourhood (Fig. 1f). At the time of formation of the solar system, the solar neighbourhood acquired the solar metallicity of 0.014 (Fig. 1e) and the reference solar iron abundance (Fig. 1f). We avoid a detailed discussion of all these various GCE trends in the present work as the essential aim in the present work is to understand the GCE of the short-lived nuclides. The general trends in the GCE models are rigorously discussed elsewhere (Sahijpal & Gupta 2013), except for the only difference that in the present case the supernova (SN II + SN Ib/c) iron yields were reduced by a factor of two compared to an identical reduction for all the iron-peaked nuclides in the earlier work.

The deduced trends in the GCE of the short-lived nuclides, $^{26}$Al, $^{36}$Cl, $^{41}$Ca, $^{53}$Mn and $^{60}$Fe, with respect to the GCE trends of their stable nuclides, $^{27}$Al, $^{35}$Cl, $^{40}$Ca, $^{55}$Mn and $^{56}$Fe, respectively, are presented in Fig. 2a and Fig. 2b for the GCE models corresponding to the pre-revised (Anders & Grevesse 1989) and the revised (Asplund *et al.* 2009) solar metallicities. In the case of the pre-revised solar metallicity model, we infer the ISM values of $3.3\times10^{-6}$ for $^{26}$Al/$^{27}$Al, $3.5\times10^{-6}$ for $^{36}$Cl/$^{35}$Cl, $6.9\times10^{-7}$ for $^{41}$Ca/$^{40}$Ca, $1.4\times10^{-4}$ for $^{53}$Mn/$^{55}$Mn and $3.5\times10^{-7}$ for $^{60}$Fe/$^{56}$Fe at the time of the formation of the solar system at ~8.5 Gyr. (Fig. 2a). The present deduced ISM values at 13 Gyr. are $2.3\times10^{-6}$ for $^{26}$Al/$^{27}$Al, $3.1\times10^{-6}$ for $^{36}$Cl/$^{35}$Cl, $5.2\times10^{-7}$ for $^{41}$Ca/$^{40}$Ca, $9.9\times10^{-5}$ for $^{53}$Mn/$^{55}$Mn and $2.7\times10^{-7}$ for $^{60}$Fe/$^{56}$Fe (Fig. 2a). The present deduced ISM value of $2.3\times10^{-6}$ for $^{26}$Al/$^{27}$Al matches perfectly with the earlier prediction (Huss & Meyer 2009; Meyer & Huss 2009) for the pre-revised solar metallicity model even though the two approaches distinctly differ from each other. However, the deduced $^{60}$Fe/$^{56}$Fe value is an order of magnitude higher than the earlier prediction (Huss & Meyer 2009; Meyer & Huss 2009).

In the case of the revised solar metallicity model (Asplund *et al.* 2009), we infer the ISM values of $5.2\times10^{-6}$ for $^{26}$Al/$^{27}$Al, $3.8\times10^{-6}$ for $^{36}$Cl/$^{35}$Cl, $4.5\times10^{-7}$ for $^{41}$Ca/$^{40}$Ca, $9.5\times10^{-5}$ for $^{53}$Mn/$^{55}$Mn and $1.5\times10^{-7}$ for $^{60}$Fe/$^{56}$Fe at the time of the formation of the solar system at ~8.5 Gyr. The present deduced ISM values at 13 Gyr. are $3.2\times10^{-6}$ for $^{26}$Al/$^{27}$Al, $3.1\times10^{-6}$ for $^{36}$Cl/$^{35}$Cl, $3.1\times10^{-7}$ for $^{41}$Ca/$^{40}$Ca, $6.3\times10^{-5}$ for $^{53}$Mn/$^{55}$Mn and $1.2\times10^{-7}$ for $^{60}$Fe/$^{56}$Fe. It should be noted that subsequent to the revision in the solar metallicity from a value of 0.02 to 0.014 (Anders & Grevesse 1989; Asplund *et al.* 2009), the GCE predictions of the revised solar metallicity model should be treated as the final predictions. The present deduced GCE value of $3.2\times10^{-6}$ for $^{26}$Al/$^{27}$Al and $1.2\times10^{-7}$ for $^{60}$Fe/$^{56}$Fe for the ISM (Fig. 2b) should be compared with the γ-ray astronomically observed (Mahoney *et al.* 1984; Diehl *et al.* 2006; Wang *et al.* 2007; Huss *et al.* 2009) values of ~$4\times10^{-6}$ for $^{26}$Al/$^{27}$Al and ~$4.8\times10^{-8}$ for $^{60}$Fe/$^{56}$Fe. The agreement in the case of $^{26}$Al/$^{27}$Al is quite good. The $^{60}$Fe/$^{56}$Fe value is also explained within a factor of two. This is considered to be reasonably well within the GCE predictions. As mentioned earlier, we have not made any attempt to suitably modify the stellar nucleosynthetic yields to explain the varied trends in the GCE predictions. A comparison between the galactic chemical evolution of the relevant stable nuclides based on the pre-revised and the revised models is presented as the ratios in Fig. 2e. It should be noted that the elemental abundance evolution of the relevant stable nuclides does not vary by more than a factor of 2.

As discussed in the previous section, on contrary to the previous works (Huss & Meyer 2009; Meyer & Huss 2009), we incorporated the stellar nucleosynthetic yields of the short-lived nuclides from supernovae SN Ia in addition to yields of SN II and SN Ib/c. In Fig. 2b, we present the GCE models with the inclusion (solid curves) and the exclusion (dashed curves) of SN Ia yields. As SN Ia are the prominent source of the intermediate nuclides, the GCE predictions for $^{53}$Mn and $^{60}$Fe are distinct for the two models. Without the inclusion of SN Ia yields, the $^{60}$Fe/$^{56}$Fe reduces to a value of ~$8.7\times10^{-8}$ for the present ISM, a value much closer to the observed ISM value. The uncertainties in SN Ia $^{60}$Fe yields could be one of the anticipated reasons for the discrepancy. As mentioned earlier, due to the unavailability of the metallicity dependent yields of SN Ia, we are bound to use to yields corresponding to the solar metallicity models (Iwamoto *et al.* 1999).

The predicted trends in the GCE of the short-lived nuclides, $^{26}$Al, $^{36}$Cl, $^{41}$Ca, $^{53}$Mn and $^{60}$Fe, as absolute mass fractions are also presented in Fig. 2c and Fig. 2d for the pre-revised (Anders & Grevesse 1989) and the revised solar metallicity (Asplund *et al.* 2009) models, respectively. As mentioned earlier, one of the major limitations of all the GCE models is the use of instant recycling mixing approximation (Matteucci 2003; Pagel 1997). According to this approximation subsequent to stellar evolution the stellar ejecta is instantaneously mixed over the entire solar annular ring, an area of ~$9.49\times10^{7}$ pc$^2$ in the present work. Thus, the predicted trends in the GCE of the short-lived nuclides are based on averaging over the entire area that is quite huge in context to the mixing time scales involved in the supernova ejecta with ISM compared to the short lifetime of the short-lived nuclides (Sahijpal & Soni 2006; Huss *et al.* 2009). A supernova shockwave associated with the ejecta interacts with ISM in a consecutive episodic manner that involves a blast wave (constant velocity), Sedov (constant energy) and snow plough (constant momentum) (Sahijpal & Soni 2006). The supernova shockwave can travel a distance of few tens of parsecs over a timescale of $10^{5-6}$ years before getting retarded and homogenized by the local ISM to random velocities < 10 km s$^{-1}$. This timescale can significantly influence the steady state contribution of the short-lived nuclides over the entire solar annular ring even though this will not significantly influence the steady state contributions of the stable nuclides except perhaps for the earliest epochs of the evolution of the galaxy (Sahijpal 2013). Even though a random series of supernovae occurring at several locations within the solar annular ring can result in a steady state evolution of the stable isotopic composition of the entire solar annular ring, yet it would be difficult to produce a steady state homogenized evolution of the short-lived nuclides, specifically the radionuclides with the lifetime < 5 Myr. (Huss *et al.* 2009). Thus, we envisage a scenario whereby the solar annular ring evolves in an un-homogenized manner as far as the short-lived nuclide abundances are concerned, with the stable nuclides being the exception. The regions within the solar annular ring associated with the evolving stellar clusters will be rich in the concentration of the short-lived nuclides, with a gradient along the radial density profile of the cluster. The regions far away from the clusters will be depleted of the short-lived nuclides. Thus, the steady state predictions of GCE for the short-lived nuclides in the Fig. 2 are based on time averaging of the stellar nucleosynthetic yields over the solar annular ring area of ~9.49×10$^7$ pc$^2$. In the vicinity of an evolving stellar cluster, the estimated ratios (Fig. 2a, b) and the absolute mass fractions (Fig. 2c, d) of the short-lived nuclides will be boosted up by a factor, '$\alpha = A_{solar-ring}/A_{stellar-cluster}$', where, $A_{solar-ring}$ is the area of the solar annular ring (~9.49×10$^7$ pc$^2$), and $A_{stellar-cluster}$ is the area associated with the evolving stellar cluster over which the stellar ejecta corresponding to the short-lived nuclides are homogenized. The factor should also accommodate the difference in the averages densities of the two associated regions. A parcel of the "local" homogenized ISM could eventually make a way into the solar system. However, the evolutionary history of the local ISM before the formation of the solar system has to be properly understood in terms of the major star forming episodes and the time intervals between them.

The giant molecular cloud (GMC) within which the solar system formed by the gravitational collapse of a single molecular cloud core probably originated over a considerable timescale after the last major stellar nucleosynthetic event in the very vicinity of the local ISM related with the formation of the solar system (Clayton 1983). In the proposed three phase mixing model (Clayton 1983), the supernova ejecta matter homogenizes with the warm and small neutral hydrogen clouds.

These small neutral hydrogen clouds eventually evolve into large cool neutral hydrogen clouds over a timescale, referred as 'T$_2$' (Clayton 1983). These large neutral hydrogen clouds eventually evolve into giant molecular clouds over a timescale, referred as 'T$_1$' (Clayton 1983). Thus, it takes a minimum time for the local ISM to evolve into a GMC for the next generation of star formation in a specific region after the last major supernovae event. This hypothesis supports the association of two stellar clusters immediately related on the temporal scale with the formation of the solar system (Sahijpal 2013). One of these stellar clusters, referred here as 'cluster-A', evolved before the formation of the GMC within which our solar system formed, and the second, referred here as 'cluster-B', that was related with the formation of the solar system. These inferences are based on the presence of the comparatively long-lived radionuclide, $^{129}$I ($\tau \sim 25$ Myr.) in the early solar system (Clayton 1983; Huss *et al.* 2009) that was perhaps produced by the stellar cluster-A.

In context to the discrete nature of the present GCE model, we can study in details the nature and the stellar nucleosynthetic contributions of the two stellar clusters, namely, cluster-A and cluster-B. The stars within the cluster-A evolve, and start contributing to the local ISM that will eventually evolve into the GMC associated with the formation of the solar system. In the absence of another major stellar cluster in its vicinity, the steady state stellar nucleosynthetic contribution of the short-lived nuclides from the GCE and the stellar cluster-A will gradually drop due to the isolation of the local ISM from the entire solar annular ring. The local ISM will eventually start evolving into GMC. The stars evolving within the cluster-A alone will locally provide the short-lived nuclides to the ISM. The gradual decline in the stellar nucleosynthetic contributions of the short-lived nuclides prior to the formation of the solar system around 4.56 Myr. are presented in Fig. 3. The shut-down in the SN II and SN Ib/c contributions occur over a timescale of ~25 Myr. after the formation of the cluster-A (Fig. 3b, d) as the massive stars evolve faster than the low and intermediate mass stars. The Fig. 3b, d were generated without any contribution from supernova SN Ia to highlight the contributions of SN II and SN Ib/c. The SN Ia contributions to the ISM from cluster-A will continue beyond ~2 Gyr. after the formation of the cluster. This is presented in Fig. 3a, c that includes the contribution of SN Ia. The stars in the low and intermediate mass range evolve slowly compared to massive stars, hence these stars in the form of SN Ia or AGB stars can contribute over long temporal scales. The $^{26}$Al contributions of the AGB stars are also presented in Fig. 3b, d. In general, apart from the stellar contributions of the short-lived nuclides from SN II and SN Ib/c, the low and intermediate mass stars can also contribute to an extent to the local ISM. However, it should be noted that as the stellar clusters evolve (Sahijpal & Gupta 2009), the stars drift out of the cluster and move out of the local ISM during tens of millions years, thereby, contributing negligibly to the local ISM. We cannot rule out the possibility of an AGB star or a star undergoing SN Ia to be eventually passing nearby the protosolar molecular cloud and contaminating the cloud with the short-lived nuclide. However, this possibility could be extremely low (Sahijpal & Soni 2006). We ignore such contributions. The contributions of the short-lived nuclides to the early solar system from the supernova SN II and SN Ib/c from the stellar cluster-A are worked out in the following. But before that we deduce the nature of the stellar cluster in terms of the number distribution of the stars. We will assume the stellar cluster-A to be identical to the stellar cluster-B that was associated with the formation of the solar system. Even though, the wide range of star forming regions indicates distinct masses of the stellar cluster (or associated stellar clusters) (Sahijpal 2013), the assumed similarity of the two massive stellar clusters will rule out the need for contributions from any other stellar cluster in the vicinity of the local ISM that has not been accounted for in our estimates.

The stellar initial mass and the number distribution of the cluster-B stars (and presumably the cluster-A stars) along with the lifetime of the various stars that will eventually undergo SN II, SN Ib/c, SN Ia and AGB stages of stellar evolution are presented in Fig. 4a. The solar system was probably formed either in the stellar cluster-B along with the other stars, or the most massive star in this stellar cluster triggered the formation of the solar system (Sahijpal & Soni 2006; Huss *et al.* 2006; Gaidos *et al.* 2009). In either case, the most massive star in the stellar cluster-B could be the main source of the short-lived nuclides. It should be noted that this specific size distribution of the cluster is based on the stellar initial mass function (eq. 1) defined by the star formation rate (SFR) at ~8.5 Gyr. (2). The vertical arrow in Fig. 4a on the stellar mass scale represents the most massive star in the cluster that will be the first to evolve, and contribute the short-lived nuclides to the early solar system (Sahijpal & Gupta 2009). We deduce the contributions of this single star to the inventories of the short-lived nuclides in the early solar system. The vertical dashed line at 8.5 Gyr. marks the beginning of the formation of the stellar cluster-B. The stellar contributions initiates after 4 Myr. by the supernova explosion of the most massive star in the cluster that contributes a value of $9\times10^{-8}$ for $^{26}Al/^{27}Al$, averaged over the entire solar annular ring (Fig. 4b). The box encloses the stellar contribution of this most massive star (84 $M_\odot$) in the stellar cluster to the early solar system inventories (Huss *et al.* 2006; Sahijpal & Soni 2006; Gaidos *et al.* 2009) if we appropriate boost-up the ratio by a factor 'α' as defined earlier. In order to explain the early $^{26}Al/^{27}Al$ value of $5\times10^{-5}$ in the solar system, with a free decay time interval of one million years between the supernova of this most massive star and the formation of the early solar system phases, a value of 1450 for 'α' would be required. This will produce the values of $3.5\times10^{-6}$ for $^{36}Cl/^{35}Cl$, $3.3\times10^{-9}$ for $^{41}Ca/^{40}Ca$, $4.2\times10^{-4}$ for $^{53}Mn/^{55}Mn$ and $4.2\times10^{-6}$ for $^{60}Fe/^{56}Fe$ for the early solar system. These values have to be compared with the initial observed estimates (Ito *et al.* 2006; Huss *et al.* 2009; Krot *et al.* 2012) of

~5×10$^{-6}$ for $^{36}$Cl/$^{35}$Cl, ~4×10$^{-9}$ for $^{41}$Ca/$^{40}$Ca, ~1×10$^{-5}$ for $^{53}$Mn/$^{55}$Mn and ~1×10$^{-6}$ for $^{60}$Fe/$^{56}$Fe for the early solar system. Except for the inferred high $^{53}$Mn, most of the radionuclide inventories of the early solar system can be explained within a factor of four by the supernova of a single massive star of the cluster-B. As pointed out earlier, the high $^{53}$Mn from a supernova can be suppressed by invoking a large cut-off mass at the time of supernova for the stellar remnant (Woosley & Weaver 1995; Huss *et al.* 2009). This has been considered as the traditional remedy against high $^{53}$Mn. Further, as mentioned earlier, the traditional approach of GCE models involves reducing the SN II and SN Ib/c yields of the iron nuclide by a factor of two. We considered this reduction for the stable isotopes of iron. However, the $^{60}$Fe yields were not reduced. If we consider a reduction in $^{60}$Fe, the single massive star is able to explain the $^{60}$Fe/$^{56}$Fe abundance in the early solar system within a factor of two.

The physical interpretation of 'α' needs some discussion in context to the differences in the average density of the ISM associated with the solar annular ring and the local ISM associated with the stellar cluster. The above deduced values of 1450 for 'α' would correspond to the stellar cluster area '$A_{stellar-cluster}$' of ~6.5×10$^4$ pc$^2$. However, the average density of the GMC within which the stellar clusters are formed could be 100-1000 times the average density of the ISM in the galaxy. Hence, we anticipate a reduced effective area of ~65-650 pc$^2$ associated with the contamination of the local ISM by a single supernova of a massive star. This range matches with the spatial dimensions of the supernova contaminated area deduced earlier by a distinct approach (see e.g., Sahijpal & Soni 2006).

The stellar contributions to the local ISM from the stellar cluster-B can continue for 25 Myr. after the explosion of the most massive star in the cluster that could have also triggered the formation of the solar system (Huss *et al.* 2006; Sahijpal & Soni 2006; Gaidos *et al.* 2009), or at least contaminated the protoplanetary disc (Sahijpal & Gupta 2009) (Fig. 4b). However, subsequent to the gravitational collapse of the protosolar molecular cloud, the subsequent stellar contributions from the supernovae of the remaining stars to the forming solar system would cease. This happens due to the increase in the central density of the collapsed protosolar molecular cloud that would probably resist any further intrusion of the matter from outside. Nonetheless, we estimate the stellar contributions of the short-lived nuclides to the local ISM exclusively from all the supernovae SN II and SN Ib/c of the stellar cluster in Fig. 4b. The vertical arrow in the Fig. 4b marks the last supernova SN II within the cluster. The exponential decay of the short-lived nuclides follows subsequently. As discussed earlier, the stellar contributions from the supernovae, SNII and SN Ib/c, from the entire stellar cluster becomes relevant in the case of the contributions of the cluster-A to the local ISM.

The stellar cluster-A can be considered to be almost identical to the cluster-B except for the difference in its formation time that could have occurred prior to the evolution of the cluster-B and the formation of the solar system. We deduce the values of the various short-lived nuclides after the last supernova SN II within the stellar cluster-A (Fig. 4b) that is marked by the vertical arrow to be $4.2\times10^{-8}$ for $^{26}Al/^{27}Al$, $1.6\times10^{-8}$ for $^{36}Cl/^{35}Cl$, $1.6\times10^{-8}$ for $^{41}Ca/^{40}Ca$, $1.0\times10^{-6}$ for $^{53}Mn/^{55}Mn$ and $2.0\times10^{-9}$ for $^{60}Fe/^{56}Fe$. We used the simple three phase model of ISM (Clayton 1983) for a value of $T_1 = T_2 = 50$ Myr. to deduce the values of $4.3\times10^{-11}$ for $^{26}Al/^{27}Al$, $2.9\times10^{-12}$ for $^{36}Cl/^{35}Cl$, $3.6\times10^{-13}$ for $^{41}Ca/^{40}Ca$, $2.0\times10^{-8}$ for $^{53}Mn/^{55}Mn$ and $2.2\times10^{-11}$ for $^{60}Fe/^{56}Fe$. These values are much less than the observed estimates based on the early solar system phases (Ito *et al.* 2006; Huss *et al.* 2009; Krot *et al.* 2012). However, if we boost-up the stellar yields by a factor of 500 for 'α', we anticipate these ratios to be $2.1\times10^{-8}$ for $^{26}Al/^{27}Al$, $1.4\times10^{-9}$ for $^{36}Cl/^{35}Cl$, $1.8\times10^{-10}$ for $^{41}Ca/^{40}Ca$, $1.0\times10^{-5}$ for $^{53}Mn/^{55}Mn$ and $1.0\times10^{-8}$ for $^{60}Fe/^{56}Fe$. Except for the $^{53}Mn/^{55}Mn$ ratio, we are not able to explain the presence of any other short-lived nuclide in the early solar system from cluster-A. If we use the timescale $T_1 = T_2 = 100$ Myr. (Clayton 1983) that is required to explain the initial abundance of the short-lived nuclide $^{129}I$ from the stellar cluster-A to the early solar system, a value of 2000 for 'α' would be required to explain the contribution of $^{53}Mn$ to the early solar system. The situation would become worse for the remaining short-lived nuclides. The contributions of $^{41}Ca$ and $^{60}Fe$, along with $^{53}Mn$ from the stellar cluster-A to the early solar system can be explained in case we use $T_1 = T_2 = 10$ Myr. and a value of 2000 for 'α'. However, this seems to be an unreasonable scenario as it is difficult to evolve the ISM to GMC over a timescale < 25 Myr. during which the supernova SN II and SN Ib/c are still going on in the cluster-A. In summary, except for $^{53}Mn$ along with $^{129}I$, we rule out the contributions of other short-lived nuclides from the stellar cluster-A to the early solar system. The other short-lived nuclides were probably produced by a single massive star as discussed in the case of the stellar cluster-B.

The essential requirement of a single local supernova contribution of $^{26}Al$ and $^{60}Fe$ to the solar system imposes a stringent constraint on the nature of the early thermal evolution of planetesimals of the planetary systems associated with other stars. As mentioned earlier, the two short-lived nuclides exclusively provide heat for the early thermal evolution and differentiation of the planetesimals and asteroids (Sahijpal *et al.* 2007) along with the wide-range of aqueous alteration (Sahijpal & Gupta 2011), and the thermal evolution of the icy planetesimals and icy satellites of the giant planets (Sahijpal 2012). In the absence of a single local supernova event associated with the formation of an exo-solar planetary system, the associated planetary system will not experience the severe early thermal evolution of its planetesimals.

## 4. Conclusions

We understand the birth of the solar system with the evolution of galaxy with the help of N-body simulation. The approach is based on understanding the evolution of the galaxy in a realistic manner by considering the origin and the evolution of stars within stellar cluster in a discrete manner. We specifically understand the nature of the stellar clusters that evolved prior to and during the formation of the solar system. We present results for the galactic chemical evolution (GCE) of the short-lived nuclides, $^{26}$Al, $^{36}$Cl, $^{41}$Ca, $^{53}$Mn and $^{60}$Fe. We found a match to the present γ-ray observed ISM values of $^{26}$Al and $^{60}$Fe with our predictions within a factor of two. However, the majority of the radionuclides in the early solar system are found to be produced by a single stellar event that was also related with the origin of the solar system. We rule out any significant contribution of the radionuclides to the early solar system from the GCE and the stellar nucleosynthetic event that probably produced $^{129}$I and $^{53}$Mn.

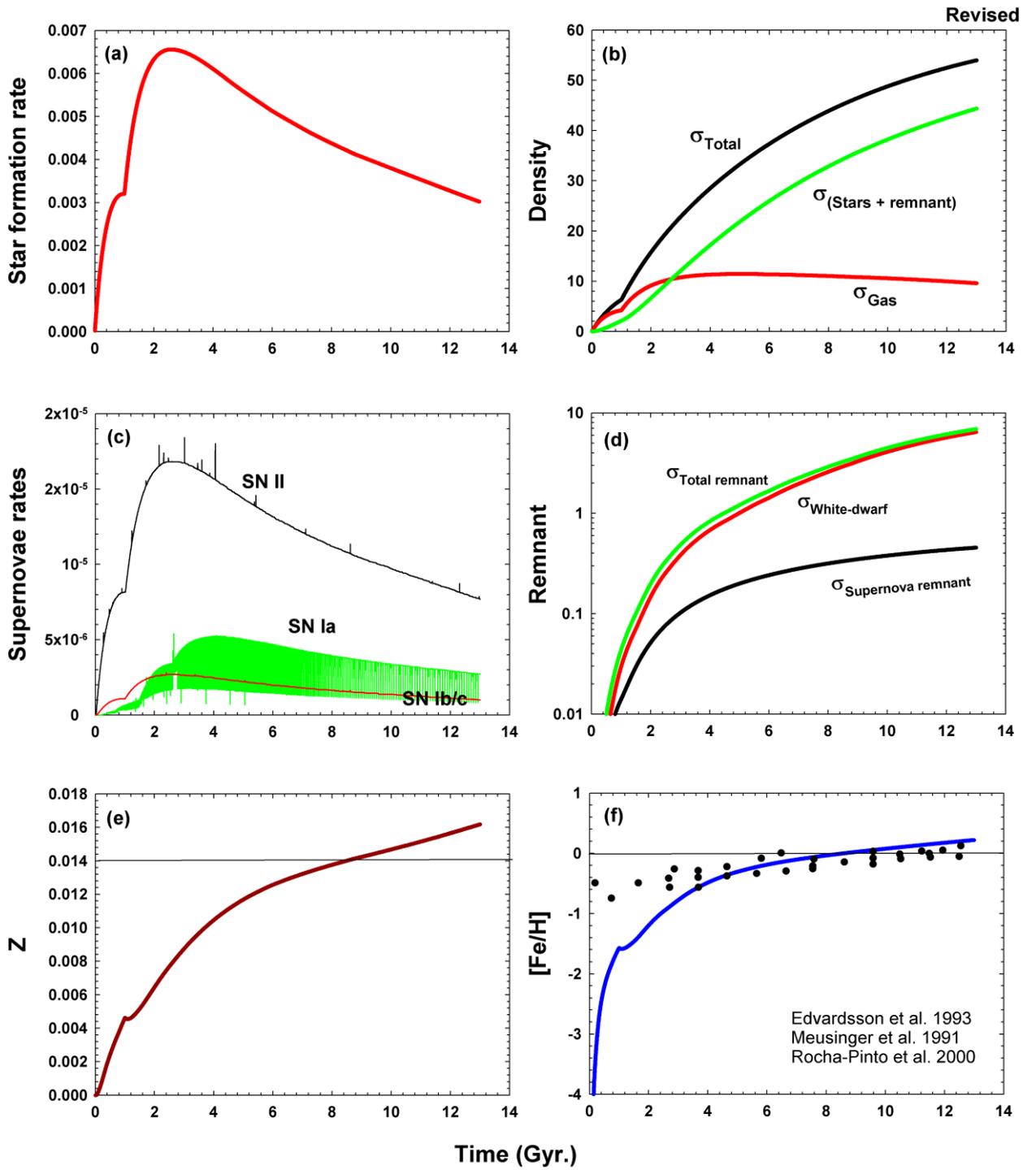

**Fig. 1.**

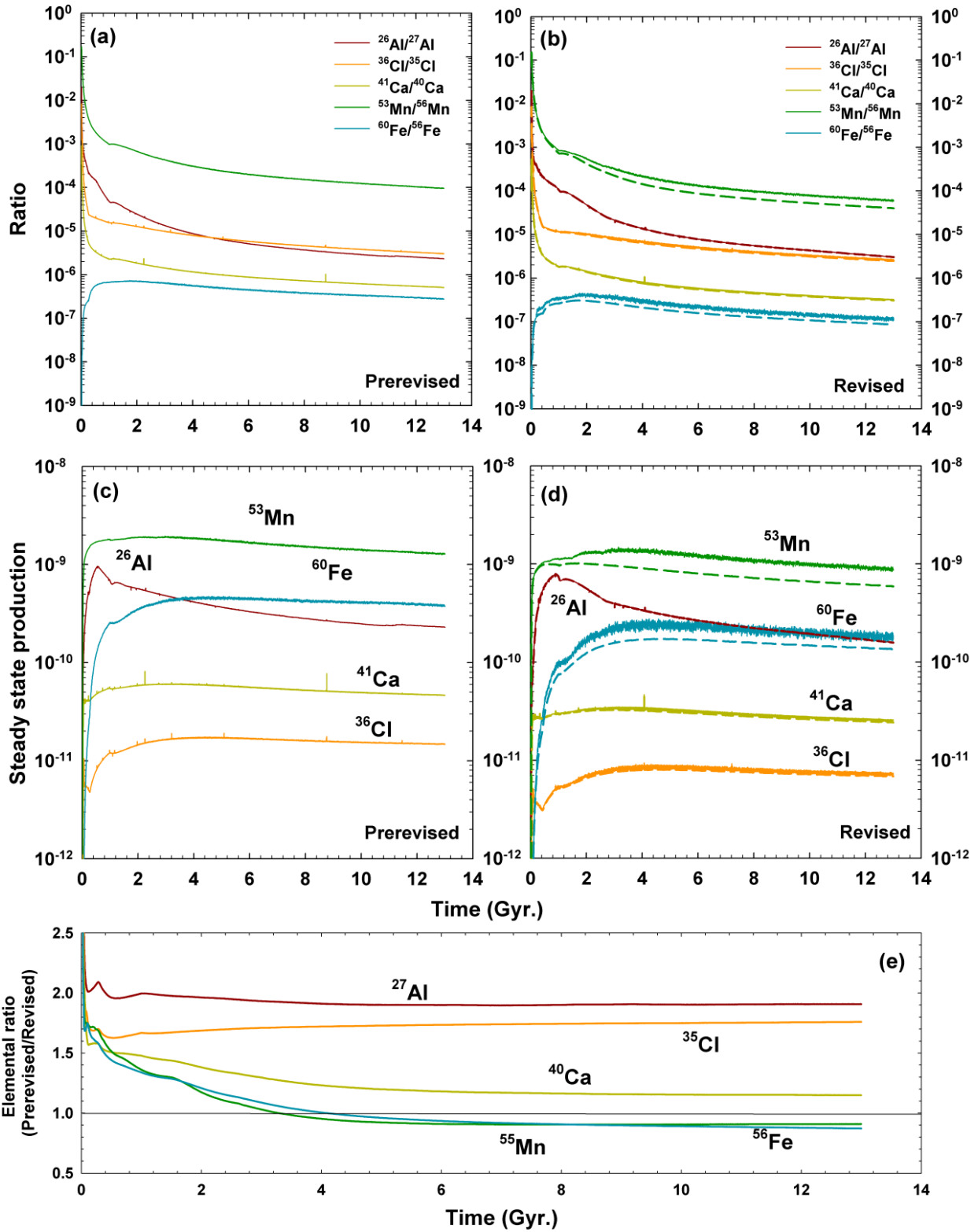

**Fig. 2.**

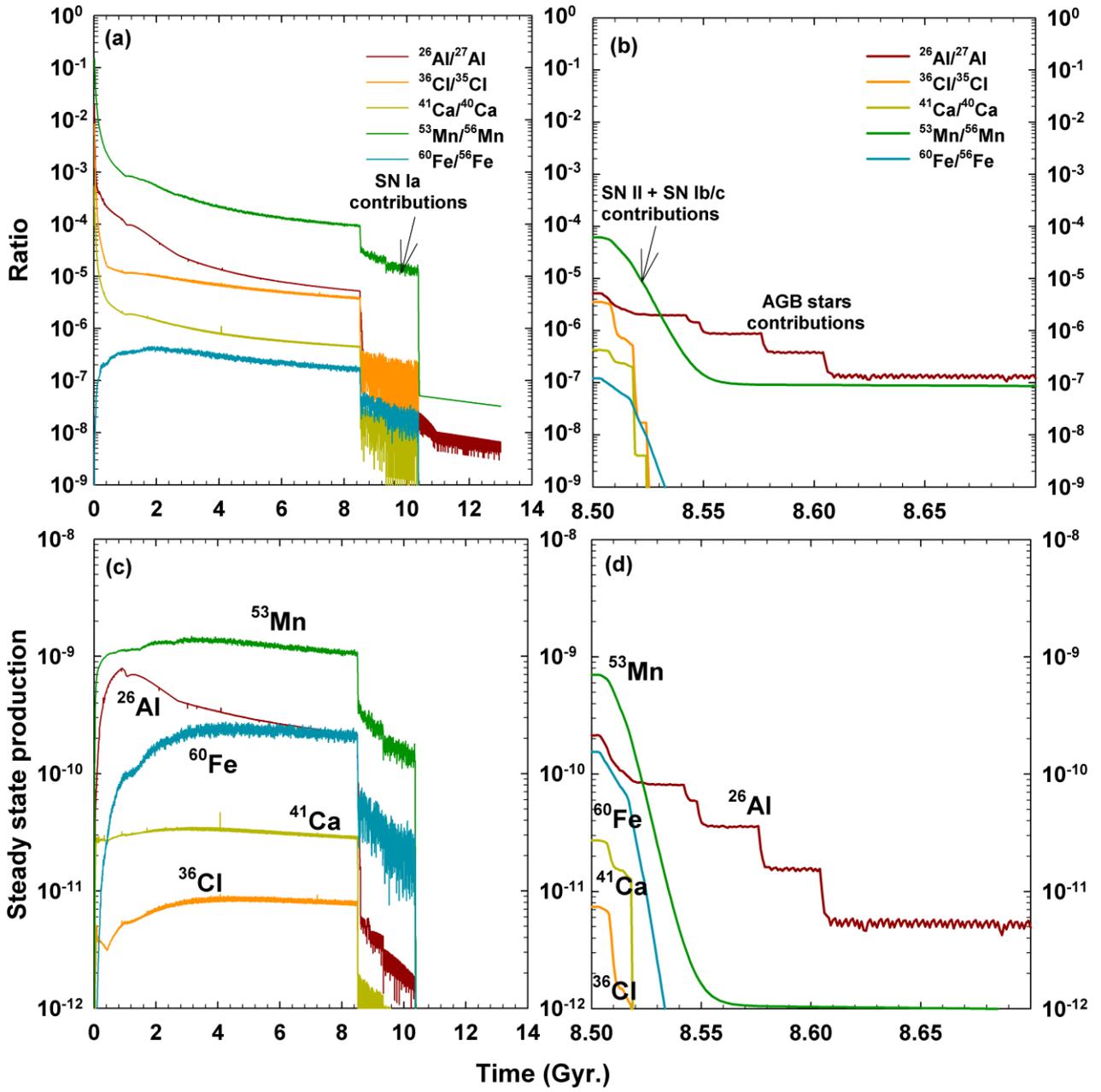

**Fig. 3.**

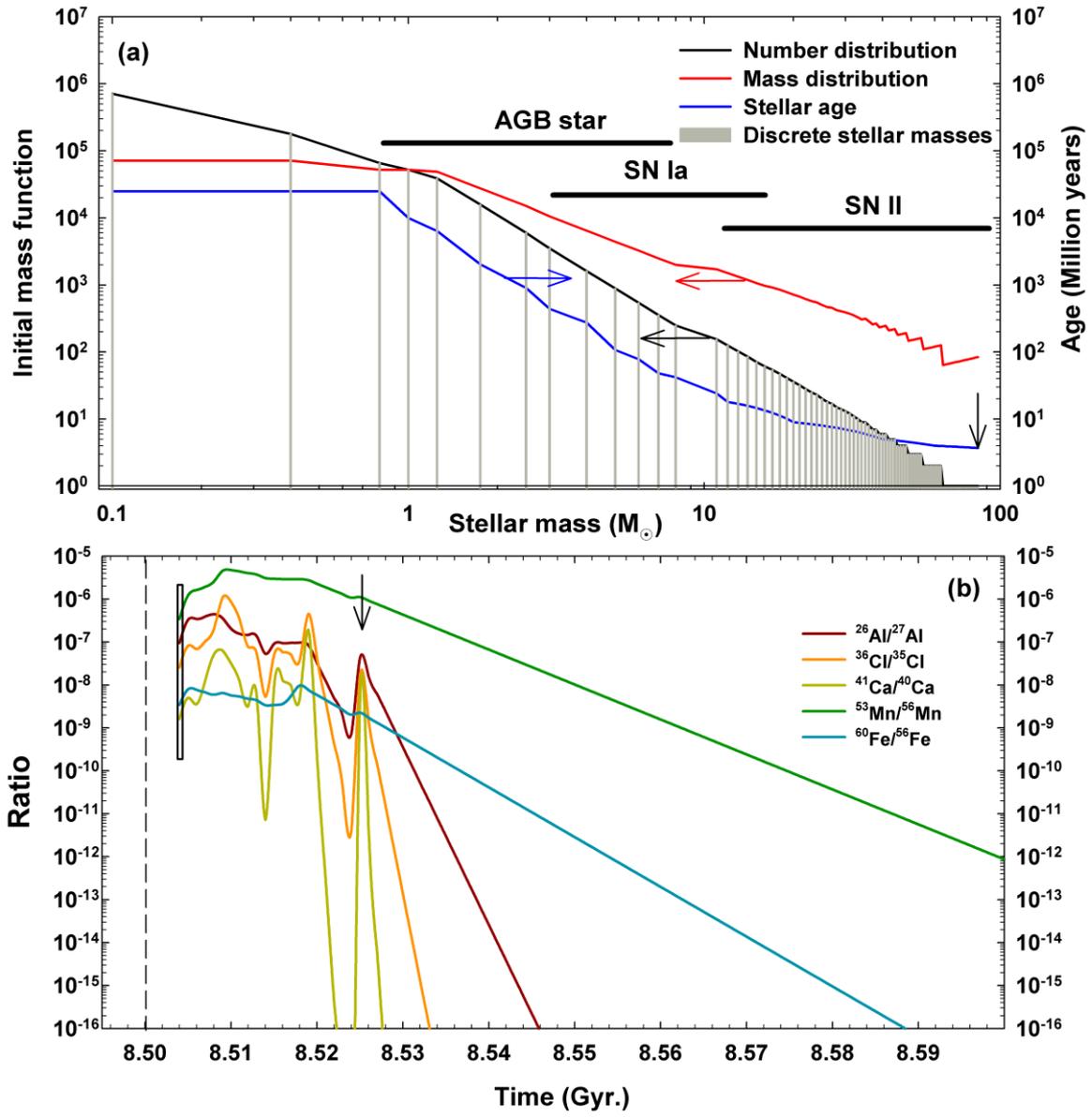

**Fig. 4.**

**Figure captions**

**Figure 1.** The deduced temporal evolution of the star formation rate (SFR; in $M_\odot$ pc$^{-2}$ Myr$^{-1}$); the total density ($\sigma_{Total}$; in $M_\odot$ pc$^{-2}$), the gas density ($\sigma_{Gas}$) and the (stars+remnant) density ($\sigma_{Stars+remnant}$); the supernovae rates (pc$^{-2}$ Myr$^{-1}$); the stellar remnant densities; the metallicity and the [Fe/H] trend in the solar neighbourhood. The metallicity, 'Z', is total mass fraction of all the elements except for hydrogen and helium. [Fe/H] = Log(Fe/H)$_{model}$ − Log(Fe/H)$_\odot$. The observational data of the F, G and K dwarf stars in the solar neighbourhood is included in order to make comparison with the deduced [Fe/H] trend. The solid horizontal lines in the metallicity and [Fe/H] trends represent the solar value that match with the trends at ~ 8.5 Giga years.

**Figure 2.** The GCE predicted trends of the short-lived nuclides, $^{26}$Al, $^{36}$Cl, $^{41}$Ca, $^{53}$Mn and $^{60}$Fe, normalized to the GCE predictions of their stable nuclides, $^{27}$Al, $^{35}$Cl, $^{40}$Ca, $^{55}$Mn and $^{56}$Fe, respectively, for the pre-revised (Anders & Grevesse 1989) and revised solar metallicity (Asplund *et al.* 2009) models. The GCE trends in the absolute mass fractions of the short-lived nuclides are also presented for the two models. The solid curves (dashed curves) represent the GCE predictions with (without) SN Ia stellar nucleoynthetic contributions of the short-lived nuclides. The various estimates should be compared with the observed initial abundances of ~5×10$^{-5}$ for $^{26}$Al/$^{27}$Al, ~5×10$^{-6}$ for $^{36}$Cl/$^{35}$Cl, ~4×10$^{-9}$ for $^{41}$Ca/$^{40}$Ca, ~1×10$^{-5}$ for $^{53}$Mn/$^{55}$Mn and ~1×10$^{-6}$ for $^{60}$Fe/$^{56}$Fe in the early solar system phases (Ito *et al.* 2006; Huss *et al.* 2009). A comparison between the galactic chemical evolution of the stable nuclides based on the pre-revised and the revised models is also presented as the ratios.

**Figure 3.** The gradual decline in the stellar nucleosynthetic contributions of the short-lived nuclides from the GCE and the stars evolving within the stellar cluster-A that were formed at a time prior to the formation of the solar system around 4.56 Myr. ago. The shut-down in the SN II and SN Ib/c contributions occur over a timescale of ~25 Myr after the formation of the cluster-A. The SN Ia contributions to the local ISM can continue for more than 2 Gyr.

**Figure 4.** The stellar initial mass and the number distribution of the cluster-B (and probably the cluster-A) stars along with the lifetime of the various stars that will eventually undergo SN II, SN Ib/c, SN Ia and AGB stages of the stellar evolution. The stellar contributions of the short-lived nuclides exclusively from the supernovae SN II and SN Ib/c of a single stellar cluster (or associated stellar clusters) are also presented. The vertical dashed line at 8.5 Gyr. marks the beginning of the formation of the stellar cluster-B. The stellar contributions initiates after 4 Myr. The box encloses the stellar contribution of the most massive star (84 $M_\odot$) in the stellar cluster. The stellar contribution can continue for 25 Myr. That is marked by a vertical arrow. The exponential decay of the short-lived nuclides follows subsequently.

## Acknowledgements

We are extremely grateful to the reviewers for suggesting several comments and suggestions that led to significant improvement of the manuscript. This work is supported by a PLANEX (ISRO) research project. We are extremely grateful to the IUCAA's associateship program.## References

Ade P. A. R. *et al.* 2013, *A&A*, arXiv:1303.5076.

Alibés, A., Labay, J., Canal, R. 2001, *A&A*, **370**, 1103.

Anders, E., Grevesse, N. 1989, *Geochim. Cosmochim. Acta*, **53**, 197.

Asplund, M., Grevesse, N., Sauval, A. J., Scott, P. 2009, *ARA&A*, **47**, 481.

Bi, S. L., Li, T. D., Li, L. H., Yang, W. M. 2011, *ApJ.*, **731**, L42.

Chang, R. X., Hou, J. L., Shu, C. G., Fu, C. Q. 1999, *A&A*, **350**, 38.

Chiappini, C., Matteucci, F., Gratton, R. 1997, *ApJ*, **477**, 765.

Clayton, D. D. 1983, *ApJ*, **268**, 381.

Clayton, D. D. 1985, in Nucleosynthesis: Challenges and New Developments, eds. W. D. Arnett and J. W. Truran (Chicago: University of Chicago Press), 65

Diehl, R., Halloin, H., Kretschmer, K., Lichti, G. G., Schönfelder B., Strong, A. W., von Kienlin, A., Wang, W., Jean, P., Knödlseder, J., Roques, J.-P., Weidenspointner, G., Schanne, S., Hartmann, D. H., Winkler, C., Wunderer, C. 2006, *Nature*, **439**, 45.

Edvardsson, B., Andersen, J., Gustafsson, B., Lambert, D. L., Nissen, P. E., Tomkin, J. 1993, *ApJ*, **275**, 101.

François, P., Matteucci, F., Cayrel, R., Spite, M., Spite, F., Chiappini, C. 2004, *A&A*, **421**, 613.

Gaidos, E., Krot, A. N., Williams, J. P., Raymond, S. N. 2009, *ApJ*, **606**, 1854.

Goswami, A., Prantzos, N. 2000. *A&A*, **359**, 191.

Huss, G. R., Meyer, B. S. 2009, 40th Lunar and Planetary Science Conference XL, id.1756.

Huss, G. R., Meyer, B. S., Srinivasan, G., Goswami, J. N., Sahijpal, S. 2009, *Geochim. Cosmochim. Acta*, **73**, 4922.

Ito, M., Nagasawa, H., Yurimoto, H. 2006, *Meteoritics & Planetary Science J.*, **41**, 1871.

Iwamoto, K., Brachwitz, F., Nomoto, K., Kishimoto, N., Umeda, H., Hix, W. R., Thielemann, F. K. 1999, *ApJS*, **125**, 439.

Karakas, A. I. 2010, *MNRAS*, **403**, 1413.
25